\documentclass[%
 reprint,
superscriptaddress,
 amsmath,amssymb,
 aps,
prl
]{revtex4-1}

\RequirePackage{snapshot}
\usepackage[english]{babel}
\usepackage{amsfonts}
\usepackage{amsmath}
\usepackage{amssymb}
\usepackage{amsthm}
\usepackage{graphicx,psfrag}
\usepackage{color}
\usepackage{dirtytalk}
\usepackage{pxfonts}
\usepackage{stmaryrd}
\usepackage{listings}
\usepackage{diagbox}
\usepackage{makecell}
\usepackage{todonotes}
\usepackage{numprint}

\usepackage{hyperref}
\usepackage{url}
\IfFileExists{date.tex}{%
  \input{date.tex}

}

\npdecimalsign{.}
\npthousandsep{\,}

\newcommand{\graphmult}{\times}

\begin{document}

\title{Percolation is Odd} 

\author{Stephan Mertens}
\email{mertens@ovgu.de}
\affiliation{Institut f\"ur Physik, Universit\"at
  Magdeburg, Postfach 4120, 39016~Magdeburg, Germany}
\affiliation{Santa Fe Institute, 1399 Hyde Park Rd., Santa Fe, NM
  87501, USA}

\author{Cristopher Moore}
\email{moore@santafe.edu}
\affiliation{Santa Fe Institute, 1399 Hyde Park Rd., Santa Fe, NM 87501, USA}

\date{\today}

\begin{abstract}
We prove a remarkable combinatorial symmetry in the number of spanning configurations in site
percolation: for a large class of lattices, the number of
spanning configurations with an odd or even number of occupied sites 
differs by $\pm 1$.  In particular, this symmetry implies that the total number of spanning configurations 
is always odd, independent of the size or shape of the lattice. The class of lattices that share this symmetry 
includes the square lattice and the hypercubic lattice in any
dimension, with a wide variety of boundary conditions.

\begin{center}
Published in \textit{Physical Review Letters} \textbf{123} 230605
(2019)
\end{center} 
 \end{abstract}



\maketitle

\section{Introduction}
\label{sec:intro}

Percolation theory started in 1941 with the work of Flory on
gelation~\cite{flory:41} and its theoretical framework was formulated
1957 by Broadbent and Hammersley~\cite{broadbent:hammersley:57}. Ever
since then, it has been a very active field of research. Despite its
maturity, most of what we know about percolation is still based on
numerical computations, for which clever algorithms
like~\cite{hoshen:kopelman:76}
and~\cite{newman:ziff:00,*newman:ziff:01} had been developed.
In contrast, exact mathematical results
are rare and mostly limited to two-dimensional systems of infinite size.
Examples include the percolation thresholds for some
planar lattices found by Sykes and Essam~\cite{sykes:essam:63} or the
celebrated formulas for the critical spanning probabilities derived
from conformal invariance by Cardy~\cite{cardy:92} and Watts~\cite{watts:96}. 

Exact results for \emph{finite} systems are even rarer,
even in two dimensions. A recent example is the discovery
of an equation that connects the average number of clusters and the
wrapping probabilities for two-dimensional percolation in periodic
lattices of any size~\cite{mertens:ziff:16}.

In this contribution, we present a new combinatorial symmetry in the number of spanning configurations in site percolation. Our result holds exactly for finite systems in all dimensions for a broad class of lattices, including the square and hypercubic lattices, and for a wide variety of boundary conditions.
In particular, this symmetry implies that the total number of spanning
configurations in these systems is odd, independent of the size
or shape of the system.

We were motivated by a startling pattern that we found
empirically by exhaustive enumerating spanning configurations on 
small instances of the square lattice. 
Let $A_{n,m}(k)$ denote the number of vertically spanning configurations
with $k$ occupied sites in a lattice with $m$ rows and $n$ columns, i.e., 
where there is a path of occupied sites connecting the 
top row to the bottom row. We can define a bivariate generating function 
\begin{equation}
  \label{eq:def-R-square}
  R_{n,m}(p,q) = \sum_{k=0}^{nm} A_{n,m}(k)\,p^k\, q^{nm-k} \, . 
\end{equation}
In particular, $R_{n,m}(-1,1)$ is the difference between the number of spanning configurations with an
even or odd number of occupied sites, 
\begin{equation}
  \label{eq:parity-sum-square}
  R_{n,m}(-1,1) = \sum_{\text{$k$ even}} A_{n,m}(k) -
  \sum_{\text{$k$ odd}} A_{n,m}(k) \, .
\end{equation}
When we computed this difference explicitly for the square lattice and
for small $n$ and $m$, we found the pattern shown in
Table~\ref{tab:square-data}. It coincides with
\begin{subequations}
  \label{eq:main}
  \begin{equation}
    \label{eq:square-parity}
    R_{n,m}(-1,1) = (-1)^{s(n,m)}
  \end{equation}
  where
  \begin{equation}
    \label{eq:square-s}
    s(n,m) = \left\lfloor \frac{m}{2} \right\rfloor n + \left\lceil\frac{m}{2}\right\rceil \, . 
  \end{equation}
\end{subequations}

\begin{table}[b]
  \centering
  \begin{ruledtabular}
  \setlength{\tabcolsep}{3pt}
  \begin{tabular}{c|rrrrrrrr}
 \diagbox{$m$}{$n$}  & 1 & 2 & 3 & 4 & 5 & 6 & 7 & 8\\\hline
1& \rule{0pt}{2.5ex} $-1$ & $-1$ & $-1$ & $-1$ & $-1$ & $-1$ & $-1$ & $-1$ \\
2& 1 & $-1$ & 1 & $-1$ & 1 & $-1$ & 1 & $-1$ \\
3& $-1$ & 1 & $-1$ & 1 & $-1$ & 1 & $-1$ & 1 \\
4& 1 & 1 & 1 & 1 & 1 & 1 & 1 & 1 \\
5& $-1$ & $-1$ & $-1$ & $-1$ & $-1$ & $-1$ & $-1$ & $-1$ \\
6& 1 & $-1$ & 1 & $-1$ & 1 & $-1$ & 1 & $-1$ \\
7& $-1$ & 1 & $-1$ & 1 & $-1$ & 1 & $-1$ & 1 \\
8& 1 & 1 & 1 & 1 & 1 & 1 & 1 & 1 
  \end{tabular}
  \end{ruledtabular}
  \caption{Values of $R_{n,m}(-1,1)$ for the square lattice with $m$ rows and $n$ columns, for either open or cylindrical boundary conditions.}
  \label{tab:square-data}
\end{table}

\begin{table}
  \centering
  \begin{ruledtabular}
   \setlength{\tabcolsep}{1pt}
  \begin{tabular}{c|rrrrrrrr}
 \diagbox{$m$}{$n$}  & 1 & 2 & 3 & 4 & 5 & 6 & 7 & 8\\\hline
 1& \rule{0pt}{2.5ex}$-1$ & $-1$ & $-1$ & $-1$ & $-1$ & $-1$ & $-1$ & $-1$ \\
2& 1 & 0 & $-1$ & 1 & 0 & $-1$ & 1 & 0 \\
3& $-1$ & 1 & $-2$ & 3 & $-5$ & 8 & $-13$ & 21 \\
4& 1 & $-1$ & $-3$ & 0 & 11 & 9 & $-32$ & $-57$ \\
5& $-1$ & 0 & $-5$ & $-1$1 & $-42$ & $-121$ & $-393$ & $-1204$ \\
6& 1 & 1 & $-8$ & $-9$ & 121 & 0 & $-1805$ & 1909 \\
7& $-1$ & $-1$ & $-13$ & 32 & $-393$ & 1805 & $-13514$ & 75135 \\
8& 1 & 0 & $-21$ & 57 & 1204 & $-1909$ & $-75135$ & 0 \\
  \end{tabular}
  \end{ruledtabular}
  \caption{Values of $R_{n,m}(-1,1)$ for the triangular lattice.}
  \label{tab:triangular-data}
\end{table}

In other lattices such as the triangular lattice, the pattern is much more complicated, see
Table~\ref{tab:triangular-data}.  The result for the square lattice is remarkable since the 
fact that $R_{n,m}(-1,1)=\pm 1$ implies that the number of spanning configurations with 
even and odd $k$ is almost identical, and in particular that the total number of
spanning configurations is always odd; to our knowledge, this was not
known before.  Last but not least, the pattern of $\pm 1$'s
given by $s(n,m)$ cries out for an explanation.

The paper is organized as follows. We start by proving
\eqref{eq:square-parity} and \eqref{eq:square-s}. We  then
generalize this result to site percolation on the hypercube
$\mathbb{Z}^d$ and, more generally, to cartesian graph products.
Then we present the most general form of our
result in terms of percolation on graph stacks.
Finally we discuss the computation of $R_{n,m}(-1,1)$ for
pairs of matching lattices.

\section{The Square Lattice}
\label{sec:square-lattice}

We compute $R_{n,m}(-1,1)$ by constructing a partial matching on the set of spanning configurations: that is, for most spanning configurations $\sigma$ we define a unique partner $\sigma'$ which is another spanning configuration, 
such that $\sigma''=\sigma$. Moreover, $\sigma$ and $\sigma'$ have opposite parity, since they differ at a single site. 
As a result, the contribution of each matched pair $(\sigma, \sigma')$ cancels in the sum~\eqref{eq:parity-sum-square}. The contribution of the remaining spanning configurations is simple enough that it can be written explicitly.

We number the rows $1,\ldots,m$ from top to bottom, and the columns $1,\ldots,n$ left to right. 
Let $\sigma$ be a spanning configuration, and suppose that row $2$ is not entirely occupied. 
Then that row has a leftmost empty site $(\ell,2)$ for some $1 \le \ell \le n$. We define $\sigma'$ by flipping the site $(\ell,1)$ in the top row immediately above this empty site, occupying it if it is unoccupied in $\sigma$ and vice versa. 

Since $\sigma$ was a spanning configuration, it has a path from the bottom row to the top row. This path arrives on the top row by passing through two adjacent occupied sites, $(x,2)$ and $(x,1)$ for some $x$. But since $(\ell,2)$ is empty, $x \ne \ell$, and this path still exists in $\sigma'$. Thus $\sigma'$ is also a spanning configuration, and clearly $\sigma''=\sigma$ as claimed. 

Let us now assume that the second row of $\sigma$ is completely occupied. Then we look for the topmost even-numbered row $r$ that is not completely occupied. We again find its leftmost empty site $(\ell,r)$, and define $\sigma'$ by flipping the site $(\ell,r-1)$ in the odd-numbered row above it. As before, $\sigma$ has a path from the bottom to row $r-1$ which passed through occupied sites $(x,r)$ and $(x,r-1)$ where $x \ne \ell$, and this path still exists in $\sigma'$. Moreover, once the path arrives on row $r-1$, it can connect immediately to row $r-2$ above it since that row is completely occupied, and from there to the top of the lattice. Thus $\sigma'$ is again a spanning configuration, and $\sigma''=\sigma$ as before.

\begin{figure}
  \centering
  \includegraphics[width=0.75\linewidth]{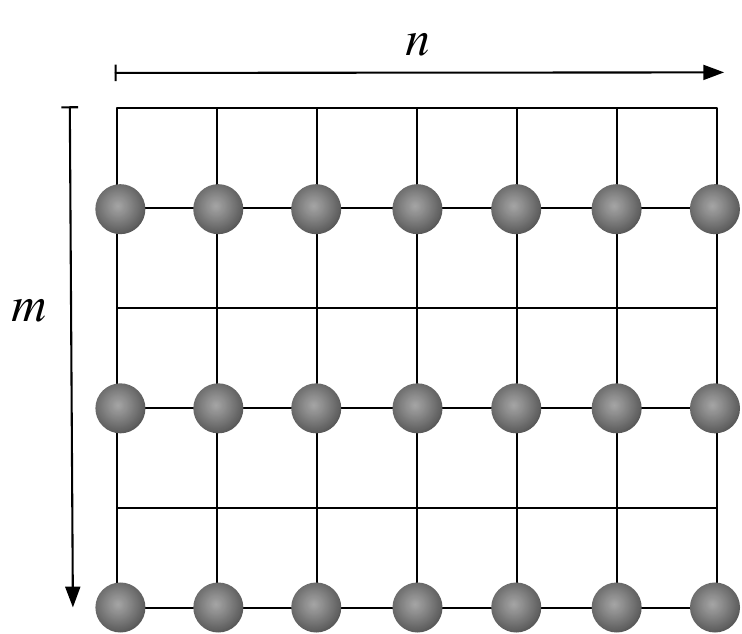}
  \caption{Only configurations in which every even numbered row is
    fully occupied need to be considered in computing \eqref{eq:parity-sum-square}. }
  \label{fig:comb}
\end{figure}

This defines an opposite-parity partner $\sigma'$ for all spanning configurations $\sigma$ except those where all even-numbered rows are fully occupied as shown in Fig.~\ref{fig:comb}. Configurations of this type are spanning configurations if and only if no odd-numbered row is completely empty. If $k_i$ denotes the number of occupied sites in row $2i-1$, the total number of occupied sites is
\[
  \left\lfloor\frac{m}{2}\right\rfloor n + \sum_{i =
    1}^{\left\lceil\frac{m}{2}\right\rceil} k_i \, ,
\]
since there are $\lfloor m/2 \rfloor$ even-numbered rows and $\lceil m/2 \rceil$ odd-numbered ones. The number of such configurations is 
\[
  {n \choose k_1} \,{n \choose k_2} \,\cdots\, {n \choose k_{\left\lceil\frac{m}{2}\right\rceil}} \, ,
\]
so~\eqref{eq:parity-sum-square} becomes
\begin{align*}
  R_{n,m}(-1,1) 
  &= (-1)^{\lfloor m/2 \rfloor n} \,
  \prod_{i=1}^{\lceil m/2 \rceil} \sum_{k_i=1}^{n} {n \choose k_i} \,(-1)^{k_i} \\
  &= (-1)^{\lfloor m/2 \rfloor n} \, \left( \sum_{k=1}^n {n \choose k} \,(-1)^k \right)^{\lceil m/2 \rceil} \\
  &= (-1)^{\lfloor m/2 \rfloor n + \lceil m/2 \rceil}
    \, ,
\end{align*}
where we used the fact that $\sum_{k=0}^n {n \choose k} (-1)^k = (1-1)^n = 0$. This proves~\eqref{eq:square-parity} and~\eqref{eq:square-s}.

\section{Graph Products and Boundary Conditions}
\label{sec:hypercubic}

The $n \times m$ square lattice with open boundary conditions can be written as the Cartesian graph product $L_n \graphmult L_m$~\cite{harary:book} where $L_n$ is the path with $n$ vertices. Let us now consider graphs of the form $G \graphmult L_m$. Each row or ``layer'' $1 \le r \le m$ is a copy of $G$, where each vertex $(v,r)$ is connected to the corresponding vertices $(v,r \pm 1)$ in the rows above and below it. For instance, if we use cylindrical boundary conditions in the horizontal direction, we obtain $C_n \graphmult L_m$ where $C_n$ is the cycle with $n$ vertices. 

What properties of $G$, if any, did we use in our proof of~\eqref{eq:main}? Suppose $r$ is the topmost even-numbered layer which is not fully occupied. Choosing the ``leftmost'' empty site $(v,r)$ in this layer can be replaced by choosing the first empty site $v$ in an arbitrary fixed ordering of the vertices of $G$. 
When we defined $\sigma'$ by flipping the site $(v,r-1)$ above this empty site, we claimed that this cannot disrupt the way the spanning path in $\sigma$ from the bottom of the lattice first arrives at layer $r-1$, since that path must go through occupied sites $(u,r)$ and $(u,r-1)$ for some $u \ne v$. As long as $G$ is connected, we can then reach any vertex in the fully occupied layer $r-2$, and from there reach the top of the lattice where $r=1$. 

Thus for any connected graph $G$, we have
\begin{equation}
  \label{eq:graphprod-parity}
  R_{G \graphmult L_m}(-1,1) = (-1)^{s(n,m)} \, ,
\end{equation}
where $n$ denotes the number of vertices of $G$ and where we define a spanning configuration as one with a path of occupied sites from the top layer $G \graphmult \{1\}$ to the bottom layer  $G \graphmult \{m\}$ of the lattice. 
This includes the case where $G$ is the $n_1 \times \cdots \times n_{d-1}$ hypercubic lattice and $n = n_1 n_2 \cdots n_{d-1}$, with any boundary conditions (open, cylindrical, toroidal, etc.). Thus~\eqref{eq:graphprod-parity} holds for the $d$-dimensional hypercubic lattice with any boundary conditions that are open in the vertical direction $n_d=m$. It also includes, for instance, the hexagonal crystal lattice in three dimensions where $G$ is a triangular lattice.

\section{Graph Stacks}

The graph product $G \graphmult L_m$ consists of $m$ identical layers, each of which is a copy of $G$. But in our proof of~\eqref{eq:main} we did not assume that the layers are identical. Hence~\eqref{eq:graphprod-parity} holds in an even more general setting, where the layers $G_1, G_2, \ldots, G_m$ are arbitrary connected graphs, each with $n$ vertices labeled $1,\ldots,n$ arbitrarily. The edges $\big( (u,r), (v,r) \big)$ within each layer $r$ coincide with the edges $(u,v)$ of $G_r$, and the edges between layers are $\big( (v,r), (v,r \pm 1) \big)$ for all $1 \le v \le n$. 

Such graphs have been considered in the study of dynamic networks, e.g.~\cite{holme:saramaki:12,ghasemian-etal}. Since there is no concept of time here, we prefer to call this construction a \emph{graph stack} and notate it 
$[G_1,G_2,\ldots, G_m]$. We again define a spanning configuration as one with a path from the top layer to the bottom layer. Then
\begin{equation}
  \label{eq:graphstack-parity}
  R_{[G_1,\ldots,G_m]}(-1,1) = (-1)^{s(n,m)} \, .
\end{equation}
This seems to be the most general case for which~\eqref{eq:main} holds. Physically, it would hold, for instance, if each layer is a connected subgraph of some $(d-1)$-lattice (perhaps with some edges removed by dilution) or if each layer has a different lattice structure entirely.

\section{Matching Lattices}
\label{sec:matching-lattices}

For lattices outside the class described in the previous section, our method to compute $R(-1,1)$ does not work. Yet for some two-dimensional lattices we can get at least partial results. For that we need the idea of the matching graph. A pair of graphs $(G,\hat{G})$ on the same vertex set is called a matching pair if both $G$ and $\hat{G}$ can be derived from an underlying planar graph $H$ in the following way. Let $F$ be the set of all faces of $H$. Then for some subset $U \subseteq F$, define $G$ (resp.\ $\hat{G}$) as a copy of $H$ with additional edges such that each face in $U$ (resp.\ $F \setminus U$) becomes a clique, i.e., a fully connected graph. Note that this definition is symmetric, so that $(G,\hat{G})$ is a matching pair if and only if $(\hat{G},G)$ is.

For example, by taking $H$ to be the square lattice and $U=\emptyset$, we see that if $G$ is the simple square lattice then its matching graph $\hat{G}$ is the square lattice with nearest and next-nearest neighbors (the Moore neighborhood). Similarly, the triangular lattice is self-matching: we can take $G$, $\hat{G}$, and $H$ all to be the triangular lattice, since its faces are already cliques. 

The crucial property of a matching pair of lattices is that a
configuration with $k$ occupied sites in an $n \times m$ lattice $G$
spans the $m$-direction if and only if the complementary configuration
consisting of the $nm-k$ empty sites does not span the $n$-direction
in the matching lattice $\hat{G}$ \cite{mertens:ziff:16}. In other words, in each
configuration, either the $k$ occupied sites span in the $m$ direction
of $G$, or the $nm-k$ empty sites span the $n$ direction on $\hat{G}$: 
\begin{equation}
  \label{eq:matching-term-by-term}
  A_{n,m}(k) + \hat{A}_{m,n}(mn-k) = {{nm} \choose k}\,,
\end{equation}
where $\hat{A}_{m,n}(k)$ denotes the number of spanning configurations
on $\hat{G}$. Multiplying this equation by $p^k q^{nm-k}$ and summing
over $k$ provides us with  
\begin{equation}
  \label{eq:matching-relation-1}
  R_{n,m}(p,q) + \hat{R}_{m,n}(q,p) = (p+q)^{nm} \, ,
\end{equation}
where $R$ and $\hat{R}$ are the generating functions~\eqref{eq:def-R-square} for spanning configurations in $G$ and $\hat{G}$ respectively. Along with the general relation
\begin{equation}
  \label{eq:argument-swap}
  R_{n,m}(p,q) = \left(\frac{q}{p}\right)^{nm} R_{n,m}(q^{-1},p^{-1}) \, ,
\end{equation}
which also also holds for $\hat{R}$, we can then write
\begin{equation}
  \label{eq:matching-relation-2}
  R_{n,m}(p,q)+
  \left(\frac{q}{p}\right)^{nm}\hat{R}_{m,n}(p^{-1},q^{-1}) = (p+q)^{nm} \, ,
\end{equation}
and in particular
\begin{equation}
  \label{eq:matching-relation-2}
  R_{n,m}(-1,1)+ (-1)^{nm}\hat{R}_{m,n}(-1,1) = 0 \, .
\end{equation}

Now for self-matching lattices like the triangular lattice, we
can take off the hat to get
\begin{equation}
  \label{eq:matching-triangular}
  R_{n,m}^{\triangle}(-1,1)+ (-1)^{nm}R_{m,n}^{\triangle}(-1,1) = 0 \, .
\end{equation}
This tells us that if $m=n$ and $mn$ is even, the balance between even and odd spanning configurations is perfect,
\begin{equation}
  \label{eq:matching-triangular-even}
  R_{n,n}^{\triangle}(-1,1) = 0 \qquad \text{($n$ even).}
\end{equation}
This pattern is visible in Table~\ref{tab:triangular-data}, where the diagonal entries with even $n$ are zero. For other values of $mn$, we only have the symmetry $R_{n,m}^{\triangle}(-1,1) = -(-1)^{nm}R_{m,n}^{\triangle}(-1,1)$. This also shows in Tab.~\ref{tab:triangular-data}, but implies no constraint on the number of configurations.

If, on the other hand, we know $R_{n,m}(-1,1)$ for a lattice $G$, we can use \eqref{eq:matching-relation-2} to compute the corresponding function for the matching lattice. Our results on the square lattice imply that for the matching lattice with the Moore neighborhood, we have
\begin{equation}
  \label{eq:nn-square}
  \hat{R}_{n,m}(-1,1) = -(-1)^{nm+s(m,n)} \, ,
\end{equation}
which also implies that the total number of spanning configurations is
odd for this lattice.

\begin{table}
  \centering
  \begin{ruledtabular}
  \setlength{\tabcolsep}{3pt}
  \begin{tabular}{r|cccccccccc}
    \diagbox{$m$}{$n$}  & 1 & 2 & 3 & 4 & 5 & 6 & 7 & 8 & 9 & 10\\\hline
 1& \rule{0pt}{2.5ex} 1 & 1 & 1 & 1 & 1 & 1 & 1 & 1 & 1 & 1 \\
 2&1 & 0 & 1 & 1 & 0 & 1 & 1 & 0 & 1 & 1 \\
 3&1 & 1 & 0 & 1 & 1 & 0 & 1 & 1 & 0 & 1 \\
 4&1 & 1 & 1 & 0 & 1 & 1 & 0 & 1 & 1 & 0 \\
 5&1 & 0 & 1 & 1 & 0 & 1 & 1 & 0 & 1 & 1 \\
 6&1 & 1 & 0 & 1 & 1 & 0 & 1 & 1 & 0 & 1 \\
 7&1 & 1 & 1 & 0 & 1 & 1 & 0 & 1 & 1 & 0 \\
 8&1 & 0 & 1 & 1 & 0 & 1 & 1 & 0 & 1 & 1 \\
 9&1 & 1 & 0 & 1 & 1 & 0 & 1 & 1 & 0 & 1 \\
 10&1 & 1 & 1 & 0 & 1 & 1 & 0 & 1 & 1 & 0 
 \end{tabular}
  \end{ruledtabular}
  \caption{Total number of spanning configurations mod 2 for the
    triangular lattice with open boundary conditions.}
  \label{tab:triangular-parity}
\end{table}

Although there is no simple pattern in $R_{n.m}^{\triangle}(-1,1)$, there appears to be an interesting pattern in the parity of the total number of spanning configurations, i.e., $R_{n.m}^{\triangle}(1,1)$. Table~\ref{tab:triangular-parity} shows this pattern for $n,m \leq 10$. For $n,m > 1$ it appears that
\begin{equation}
  \label{eq:triangular-parity}
  R_{n,m}^{\triangle}(1,1)\bmod 2 = \begin{cases}
    0 & \text{if $n-m = 0 \bmod 3$,} \\
    1 & \text{otherwise.}
  \end{cases}
\end{equation}
Proving this equation would require a different approach than the one
used in this paper, but the self-matching property of the trangular
lattice gives some information. Since the total number of spanning
configurations is given by $R_{n,m}(1,1)$, we can use
\eqref{eq:matching-relation-2} and the self-matching property to get
\begin{equation}
  \label{eq:triangular-total-symmetry}
  R_{n,m}^{\triangle}(1,1) + R_{m,n}^{\triangle}(1,1) = 2^{nm}\,.
\end{equation}
This tells us, that for $n=m$ exactly half of the $2^{n^2}$
configurations are spanning, and that the parity matrix
$R_{n,m}^{\triangle}(1,1)\bmod 2$ is symmetric. This is not sufficient
to prove \eqref{eq:triangular-parity}. We leave this as an open
question.

\section{Conclusions}

In the theory of computational complexity~\cite{noc}, counting problems---specifically, counting solutions to a problem where each solution can be verified in polynomial time---constitute the complexity class \#P. Computing the coefficients of the generating function $R_G(p,q)$ for a general graph $G$ falls into this class. For general graphs, the generating function $R_G(p,q)$ is also known as the reliability polynomial, and its computation has been shown to be \#P-complete~\cite{valiant:79b}, meaning that it is among the hardest counting problems in this class. As a result, we believe that any algorithm needs to perform some kind of explicit enumeration, and therefore requires exponential time. This is true even when $G$ is restricted to planar graphs with bounded degree~\cite{provan:86}. 

Computing a generating function like $R_G(p,q)$ at arbitrary values of $p$ and $q$ is just as hard as computing its coefficients, since we can recover its coefficients by interpolation. However, many generating functions that are \#P-complete to compute in general can be efficiently computed at specific points, such as the Tutte polynomial~\cite{jaeger:vertigan:welsh:90}. Similarly, computing the parity of a counting problem may or may not be difficult. The permanent of a matrix with $\{0,1\}$ entries is \#P-complete~\cite{valiant:79}, but its parity is easy since the permanent and determinant are equivalent modulo $2$. On the other hand, solving general counting problems mod $2$ is probably very difficult, since a polynomial-time algorithm with access to an oracle for such problems can solve any problem in the polynomial hierarchy, including problems well beyond NP-completeness~\cite{toda:91}.

In this contribution we have shown that for a large class of lattices we can compute $R_G(-1,1)$, i.e., the difference between the number of spanning configurations with an even or odd number of occupied sites. In particular, we have shown that these sets of configurations are almost perfectly balanced. In addition to implying that the parity of the total number of spanning configurations is odd, this almost-perfect balance can be used as a sanity check for enumeration algorithms that compute the numbers $A_{n,m}(k)$ for small values on $n$ and $m$.

We note that since our preprint appeared, alternate proofs were given
by Appert-Rolland and Hilhorst~\cite{appert-rolland:hilhorst:19} that
the number of spanning configurations is odd (though without the
symmetry between odd and even $k$) and by Karzes using a transfer
matrix approach (unpublished).

\acknowledgements{ We are grateful to Bob Ziff and Alex Russell for
  their input, and to Andrea Wulf and Chaco Wolpert for their
  support. S.M. thanks the Santa Fe Institute, Tracy Conrad, and
  Rosemary Moore for their hospitality. C. M. is supported by NSF
  Grant No. IIS-1838251.
}
\bibliography{parity}

\end{document}